\documentclass[PRL,twocolumn,superscriptaddress,showpacs,preprintnumbers,floats]{revtex4}
\usepackage{amssymb}
\usepackage{graphicx}

\begin{document}

\title{Connectivity of edge and surface states in topological insulators}
\author{Yongjin Jiang}
\affiliation{Center for Statistical and Theoretical Condensed
Matter Physics, and Department of Physics, Zhejiang Normal
University, Jinhua 321004, People's Republic of China}
\affiliation{Department of Physics, Purdue University, West
Lafayette, Indiana 47907, U.S.A.}
\author{Feng Lu}
 \affiliation{Center for Statistical and Theoretical
Condensed Matter Physics, and Department of Physics, Zhejiang
Normal University, Jinhua 321004, People's Republic of China}

\author{Feng Zhai}
\affiliation{Center for Statistical and Theoretical Condensed
Matter Physics, and Department of Physics, Zhejiang Normal
University, Jinhua 321004, People's Republic of China}

\author{Tony Low}
\affiliation{IBM T.J. Watson Research Center, Yorktown Heights,
NewYork 10598, USA}

\author{Jiangping Hu}
\affiliation{Department of Physics, Purdue University, West
Lafayette, IN 47907, U.S.A.}

\date{\today}

\begin{abstract}
The edge states of a two-dimensional quantum spin Hall (QSH)
insulator form a one-dimensional helical metal which is
responsible for the transport property of the QSH insulator.
Conceptually, such a one-dimensional helical metal can be attached
to any \emph{scattering region} as the usual metallic leads. We
study the analytical property of the scattering matrix for such a
conceptual multiterminal scattering problem in the presence of
time reversal invariance. As a result, several theorems on the
connectivity property of helical edge states in two-dimensional
QSH systems as well as surface states of three-dimensional
topological insulators are obtained.
 Without addressing real model details, these
 theorems, which are phenomenologically obtained,  emphasize the general connectivity property of
topological edge/surface states from the mere time reversal
symmetry restriction.

\end{abstract}

 \maketitle

\section{introduction}

 Time reversal symmetry (TRS) has profound and sometimes
mysterious consequences in quantum physics. Recently, in the
frontier of condensed matter physics, the exciting development of
two-dimensional (2D) and three-dimensional (3D) topological
insulators (TIs)\cite{Qi_2010,kane} marks a new depth of our
understanding of TR in the quantum exploration of the material
world. The TI materials have strong spin-orbital coupling (SOC)
while maintaining TRS. Prominently, these materials are
characterized by a nontrivial band structure with gapped bulk
spectrum while their edge excitations are gapless. Among divergent
research activities in this field, the theoretical proposal of 2D
quantum spin Hall (QSH)
insulators\cite{Kane1,Kane2,Bernevig_2006_PRL,Bernevig_2006}
 and the
experimental confirmation\cite{Konig_2007,A.Roth}
are of core importance to the whole field. For our purpose,  we
would like to point out especially that the nonlocal transport
measurement has confirmed that the transport property of a 2D QSH
insulator is dominated by helical edge states near its
edges\cite{A.Roth}.

Different from the traditional integer quantum Hall system, where
TRS is broken by a magnetic field and the edge states are chiral,
the edge states in the QSH system are helical, which are composed
of pairs of counter-propagating modes with opposite spin
polarizations. For each pair, the two branches of states transform
into each other under a TR transformation. Due to their conducting
property, such helical edge states are called the helical
liquid\cite{Congjun Wu}. In the 2D QSH phase, the helical states
are localized near edges and separated spatially by the gapped
bulk region. In this paper, we dub the phrase \emph{helical metal}
to refer to the one-dimensional (1D) metal for which the
low-energy dispersion is characterized by a pair of helical edge
states. Such 1D helical metals are isolated from each other by a
macroscopic distance (e.g., the width of the Hall bar sample).

The Landauer-Buttiker theory (LBT) is one of the
most important frameworks
 for analyzing the transport property of mesoscopic
systems\cite{Datta}. In the LBT, the transport process
 is treated as a quantum scattering problem where the connection
between the carrier reservoir (electrical contacts) and the
mesoscopic system (scattering region) is modeled as semi-infinite
metallic leads. The central quantity in LBT is the scattering
matrix, which can be different by using different leads. In
practice, the metallic leads can be described by an arbitrary
single-particle Hamiltonian with some propagating modes for a
given Fermi energy.

In this paper, we will conceptually use the fore-mentioned helical
metals as metallic leads and attach them to the central scattering
region.
For such a conceptual scattering problem, we find that TRS imposes
a strong restriction on the form of the scattering matrix. Then,
the condition of a physically realizable scattering problem is
obtained in \texttt{Theorem A}. This restriction has profound
consequences on the connectivity property of edge states. Based on
it, we discuss the connectivity properties for edge states in the
2D QSH system (embodied in \texttt{Theorem B}) as well as the
surface Dirac cone in the 3D TI (\texttt{Theorem C}). Several
discussions for these theorems are provided.


\section{scattering matrix with helical metal as leads}
To begin with, let us consider a system with TRS which is attached
with two half-infinite helical metals at its left and right sides.
At energy $E$, the left lead has two eigenstates, denoted as $|1s
\rangle_{L}$ and $|\bar{1}\bar{s} \rangle_{L}$, while the right
lead has eigenstates $|\bar{1}s \rangle_{R}$ and $|1\bar{s}
\rangle_{R}$, where 1 refers to right-moving and $\bar{1}$ to
left-moving and $s$,$\bar{s}$ are two spin polarizations with
respect to some spin quantization axis.  Two such states form a
Kramers's pair in each lead, so that they change to each other
under time reversal operation $T$. By a proper energy-dependent
U(1) gauge fixing, the two states satisfy
\begin{eqnarray}
 T|1s
\rangle_{L}&=&|\bar{1}\bar{s} \rangle_{L},T|\bar{1}\bar{s}
\rangle_{L}=-|1s \rangle_{L} \nonumber\\
 T|\bar{1}s\rangle_{R}&=&|1\bar{s} \rangle_{R},T|1\bar{s}
\rangle_{R}=-|\bar{1}s \rangle_{R} \label{eq:1}
\end{eqnarray}
Through the above equation, $T^2=-1$ is respected for spin-1/2
particles. In general, the spin quantization axis as well as wave
vectors for the eigenstates is different for the two leads.

For the scattering problem, we generally assume
the wave functions on the two leads as:
\begin{eqnarray}
|\psi\rangle_{L}&=&\phi^{in}_{Ls}|1s
\rangle_{L}+\phi^{out}_{L\bar{s}}|\bar{1}\bar{s} \rangle_{L}\nonumber\\
|\psi\rangle_{R}&=&\phi^{out}_{R\bar{s}}|1\bar{s}
\rangle_{R}+\phi^{in}_{Rs}|\bar{1}s \rangle_{R} \label{eq:2}
\end{eqnarray}
in which $\phi^{in}_{Ls}$, $\phi^{in}_{Rs}$ are
incident wave amplitudes and
$\phi^{out}_{L\bar{s}}$, $\phi^{out}_{R\bar{s}}$
are outgoing amplitudes. It is convenient to
introduce the incident wave vector
$a=(\phi^{in}_{Ls},\phi^{in}_{Rs})^T$ and the
outgoing wave vector
$b=(\phi^{out}_{L\bar{s}},\phi^{out}_{R\bar{s}})^T$.
In the standard scattering problem, the
scattering matrix $S$ can be defined so that we
have $b=Sa$. Due to particle number conservation,
$S$ must be a unitary matrix with a proper
normalization\cite{Datta}, so that $S^{\dag}S=1$,
from which we can get $a^*=S^{T}b^*$. Now let us
consider the consequence of TRS. From
Eq.~(\ref{eq:1}), we know that under $T$:
$a\Rightarrow a'=Ta=-b^{*}$, $b\Rightarrow
b'=Tb=a^{*}$ and $S'=TST^{-1}=S$. Thus we have
$a^*=-Sb^*$. Putting these pieces together, we
get the following antisymmetry condition for the
scattering matrix:
\begin{equation}
S^{T}=-S. \label{eq:3}
\end{equation}
This is a strong restriction on the form of the scattering matrix
due to the existence of TRS for the whole system. It turns out
that a lot of interesting results can be derived from this
property. Now let us discuss them as follows.

First, from the antisymmetry property and the unitarity condition,
we can conclude that
 $S=\left(%
\begin{array}{cc}
  0 & e^{i\phi} \\
 -e^{i\phi}& 0 \\
\end{array}%
\right).$ Here $\phi$ is a real phase. This
result indicates that near the edge of a QSH
system, the helical state has no back scattering
without $T$-breaking barrier or impurities, which
is a well-known property\cite{Qi_2010}. We may
interpret this result in the connectivity
property of helical states: any $T$-invariant
barrier or impurities can not break the
connectivity of helical states.

Second, following the same procedure, the derivation of
Eq.~(\ref{eq:3}) can be easily extended to the case with an
arbitrary number, say $n$, of helical metal leads attached to the
central region, with each lead again characterized by one pair of
helical states. As a mathematical fact, the determinant of an
antisymmetric matrix of odd dimension is zero, i.e., $\det(S)=0$
when $n$ is odd. Consequently, for that case, $S$ can not be a
unitary matrix, which is one of the very assumptions that leads us
to Eq.~(\ref{eq:3}). To note, the unitarity property of the
scattering matrix is the direct consequence of the conservation
law of the particle number. What is the meaning of such a logical
contradiction? In the above conceptual scattering problem model,
we have assumed that $n$ helical metal leads are attached to the
central region and thus form a standard multiterminal scattering
problem. While the helical states near two edges of a QSH system
are separated spatially by the gapped bulk region, which
encouraged us to coin the concept of "helical metal" for the
conducting edge, they are actually topologically correlated. The
above logical contradiction means the impossibility of a
reasonable scattering matrix in a conceptual scattering problem
with an odd number of helical leads. Based on these analysis, we
can state the following theorem:

 \texttt{Theorem A} :\emph{ In a physical
scattering problem with TRS, any central region allows only an
even number of helical metals (each with a single pair of helical
states) as conduction leads attached upon it.}

It can also be straightforwardly shown that the possible existence
of any normal leads (which, by definition, must be composed of an
even number of helical pair states) in the conceptual scattering
problem does not change the statement in \texttt{Theorem A}. It is
noteworthy that a related theorem was given by Wu \emph{et
al.}\cite{Congjun Wu}, in which they proved that the helical metal
can not be realized in 1D lattice models with TRS, which is termed
a no-go theorem. It turns out that this no-go theorem is a direct
consequence of \texttt{Theorem A}: if we can construct a 1D model
to be a helical metal, then such a helical metal will become a
physical system by itself (instead of being an edge subsystem of
another system). So, we can use it as a single independent lead to
attach to a TRS central region, which contradicts \texttt{Theorem
A}. Physically, \texttt{Theorem A} and the no-go theorem have a
common origin, i.e., both of them are a direct consequence of TRS.
However, being expressed in the scattering language,
\texttt{Theorem A} is more flexible to use. Since only the parity
of the number of pairs of helical states matters, from now on, we
can loosen the previous definition of helical metal such that
their energy spectrums are characterized by any odd number(instead
of just one) of helical states. Through \texttt{Theorem A}, we
will be able to prove several rigorous properties in the
following.

\begin{figure}[tbp]
\centering
\scalebox{0.4}[0.4]{\includegraphics*[viewport=120 70 650 440]{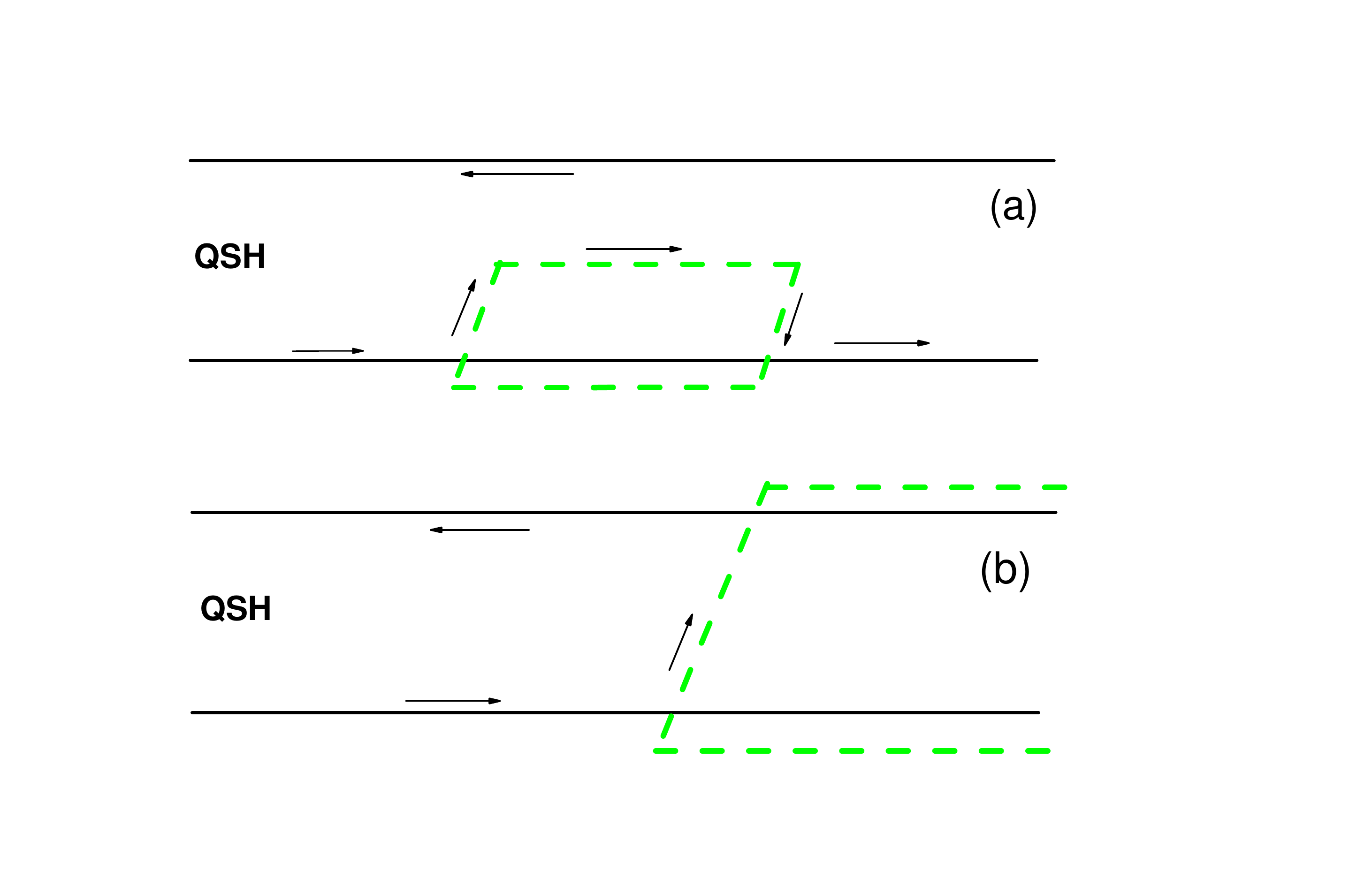}}
\caption{(Color online) 2D QSH states and helical edge states in a
stripe geometry. (Black arrows denote propagating direction for
one state of the edge helical pair). The band insulator (bordered
by green dashed lines) is put on the QSH strip (a) on one edge and
(b) bridging two edges.
}
\label{Fig1}
\end{figure}

\section{connectivity property of edge/surface states of the 2D/3D topological insulators }
 First, let us consider the 2D QSH system. We will prove that all
helical metals should be connected and form a closed loop in a
finite system. This result is quite easy to prove starting from
\texttt{Theorem A}. Suppose there is a section of line $ab$ (with
two ends $a$ and $b$),which is made of helical metal and one of
its two ends, say $a$, does not belong to, or, is not connected
with, any other section of line which is also a helical metal. If
such an end $a$ is chosen to be the central region of the
conceptual scattering problem, then, there is only one helical
metal attached to $a$, which is contrary to \texttt{Theorem A}.
Therefore, we can conclude that this can not physically happen. In
Fig.~1(a), we show that the helical state at one edge of the QSH
system can circumvent any barrier with TRS and transmit to the
other side. If the barrier is chosen to be an insulator under
which a forbidden region is defined, the perfect transmission
happens by a new helical edge formed around the boundary of the
barrier [as shown in Fig.~1(a)].
Furthermore, if the barrier is large enough to bridge the two
edges as shown in Fig.~1(b), the helical states from one edge will
connect to the other side through the interface between the QSH
system and the barrier. From the above discussion, it can be known
clearly that there are always helical states near the boundary
along which we cut the QSH system. We can summarize this result in
the following theorem:

 \texttt{Theorem B}:\emph{Along any edge of a 2D QSH system
 or a boundary between a 2D TI and a band insulator, there is
always an odd number of pairs of helical states.}


Particularly, \texttt{Theorem B} implies the $Z_2$
connectivity(evenness/oddness of the number pair of helical
states) property between the adjacent edges of any finite 2D QSH
sample. For further illustration, let us address the graphene
stripe with an intrinsic SOC\cite{Kane1,Kane2} as an example. It
is well known that a graphene stripe with zigzag edges has
zero-energy flat bands near the zigzag edges\cite{flatzigzag}. In
the presence of an intrinsic SOC, a bulk gap will open and these
flat bands will evolve into helical edge states\cite{Kane1,Kane2}.
However, for a graphene stripe with armchair edges without any
SOC, there are no such flat bands. By mere expectation through a
continuity consideration, we may anticipate that there are no
helical edge states with the inclusion of an intrinsic SOC.
However, from \texttt{Theorem B}, we can predict helical edge
states also exist near armchair edges, which is in consistent with
numerical results\cite{Ruibao Tao}.

The famous bulk-edge correspondence theorem in the quantum Hall
effect states that the nontrivial topological band structure will
ensure the existence of conducting edge modes along system edges,
with the number of edge modes determined by the topological Chern
number of the filled bands\cite{Hatsugai_PRB_1993}. This theorem
is generalized with limited success to the QSH case\cite{Qi_2006},
which involves sophisticated topological analysis.
 \texttt{Theorem B} is a direct consequence of the much sought-for
 bulk-edge correspondence theorem for
2D QSH systems, where only the parity of the
number of edge states is important.
Our approach leading to \texttt{Theorem B}, though somewhat
phenomenological (see discussion below), is model-independent and
general. Furthermore, it can be extended to 3D TIs as discussed
below.

\begin{figure}[tbp]
\centering
\scalebox{0.4}[0.4]{\includegraphics*[viewport=120 170 650 400]{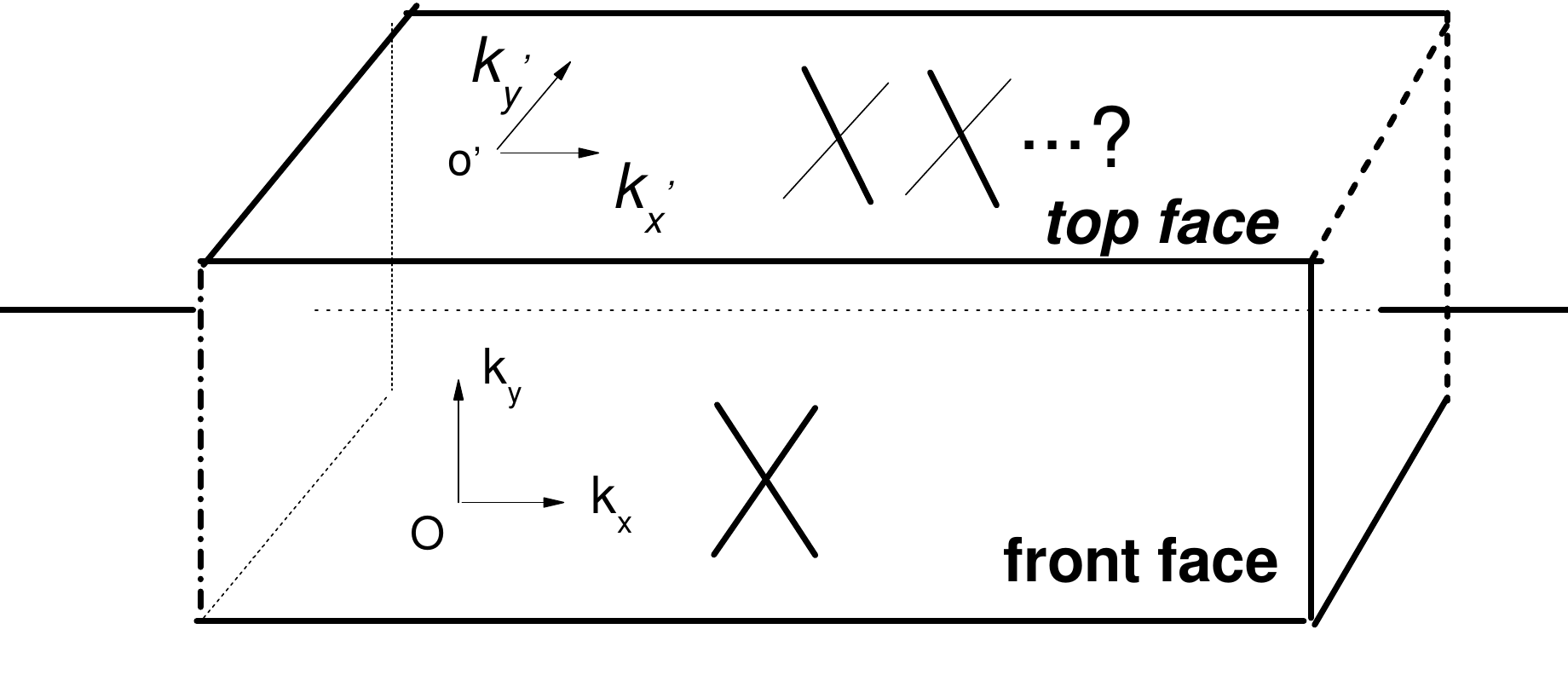}}
\caption{3D TI in a pipe geometry with translationally invariant
cross section. The local coordination for the wave vector is drawn
for the front and the top faces. If the low-energy Hamiltonian of
the front face is characterized by a single Dirac Cone, how many
Dirac cones are on the top surface?
}
\label{Fig2}
\end{figure}

 Now let us turn to the 3D TI, which has also
been firmly established both theoretically and
experimentally\cite{LFu,LFu2007PRB,HZhang,DHsieh,DHsieh2009}. As
depicted in Fig.~\ref{Fig2}, we will consider an infinitely long
rectangular column composed of a 3D TI. The system is
translationally invariant in the $x$ direction. The size of its
rectangular cross section is of macroscopic scale so that the
quantum confinement effect can be neglected in our discussion.
  As is well known, a 3D TI is
characterized by bulk gap and mid-gap surface excitations. Now,
let us assume that the front face is characterized by a single
surface Dirac cone (e.g., in $Bi_{2}Te_{3}$) around the center of
the Brillouin Zone ($\Gamma$ point). In the following, we shall
prove that the top face is also characterized by an odd number of
Dirac cones. Let us denote the wave vector $k_x,k_y$ for the front
face and $k_x',k_y'$ for the top face (as the local coordinate
frames drawn in Fig.~2). Since the system is translationally
invariant in $x$ direction, $k_x$ and $k_x'$ are good quantum
numbers. On the other hand, we can linearly combine the degenerate
states with different $k_y$ (or $k_y'$) and use the proper
boundary condition to obtain the wave function around the
perimeter of the cross section. The geometrical edge between the
front and the top face can be regarded as a scattering region for
each $k_x$-fixed subspace, thus we have a conceptual scattering
problem. At a general $E$, we have surface states,
$|\psi(k_x,k_{y})\rangle_f$ and $|\psi(k_x,-k_{y})\rangle_f$ for
the front face. On the other hand, according to TRS,
$|\psi(k_x,k_{y})\rangle_f$ and $|\psi(-k_x,-k_{y})\rangle_f$ form
a Kramers's pair. Picking up the $k_x=0$ case,
$|\psi(0,k_{y})\rangle_f$ and $|\psi(0,-k_{y})\rangle_f$ form a
Kramers's pair (or a helical metal by using the foregoing
terminology) so that they cannot scatter into each other in a TRS
scattering process [all diagonal elements of the antisymmetric
scattering matrix are zero, see Eq.~(3)]. By \texttt{Theorem A},
there must be an odd number of Kramers's pair states on the top
surface. For simplicity, we consider the case where there is just
one pair of Kramers's states $|\psi(0,k_{y}')\rangle_t$ and
$|\psi(0,-k_{y}')\rangle_t$ (later we will generalize this result
to the case of an odd number of Kramers's pairs). Perfect
tunneling occurs from $|\psi(0,k_{y})\rangle_f$ to
$|\psi(0,k_{y}')\rangle_t$. This is similar to Klein tunneling
phenomena for relativistic particles\cite{klein,katsnelson}. It
has important implications. At the edge position, which is common
to the front face and top face, the wave functions
$|\psi(0,k_{y})\rangle_f$ should be equal to
$|\psi(0,k_{y}')\rangle_t$. However, as we assumed before, the
quantum confinement effect is neglected and the eigenstates
$|\psi(0,k_{y})\rangle_f$ are nothing but plane wave spinor
eigenstates for an infinite plane.
This property will be useful when one tries to write an effective
continuum Hamiltonian for a particular surface for such one Dirac
cone case.

When $k_x$ is away from but still near to $0$,
$|\psi(k_x,k_{y})\rangle_f$ can be scattered into
$|\psi(k_x,-k_{y})\rangle_f$ with a finite probability, since they
are not TR pairs. At the same time, due to the continuity of the
physical property with respect to the parameters, there will be
degenerate states $|\psi(k_x,k_{y_1}')\rangle_t$ and
$|\psi(k_x,-k_{y_2}')\rangle_t$ ($k_{y_1}'$ may be different from
$k_{y_2}'$ in general) on the top surface with their TR partners.
These two states, together with the two states on the front
surface, constitute an even number of helical states for the
scattering problem (their TR partners being grouped into another
subspace characterized by -$k_x$).
The boundary condition is that the wave function should be
continuous at the edge, which gives two complex equations. Thus,
two unknown coefficients for $|\psi(k_x,k_{y1}')\rangle_t$ and
$|\psi(k_x,-k_{y})\rangle_f$ in the scattering problem can be
solved exactly. Perfect reflection from
$|\psi(k_x,k_{y})\rangle_f$ to $|\psi(k_x,-k_{y})\rangle_f$ may
occur but only accidentally.  On the other hand, if
$|\psi(k_x,-k_{y_2}'\rangle_t$ is the incident wave from top
surface side onto the edge, then the reflection coefficient
$|\psi(k_x,k_{y1}')\rangle_t$ and transmission coefficient
$|\psi(k_x,-k_{y})\rangle_f$ can be solved as well. Thus, the
scattering matrix can be determined exactly. By tuning $k_x$
continually, as $|\psi(k_x,k_{y})\rangle_f$ and
$|\psi(k_x,-k_{y})\rangle_f$ cover the whole Dirac cone on the
front surface once, the corresponding states
$|\psi(k_x,k_{y1}')\rangle_t$ and $|\psi(k_x,-k_{y_2}'\rangle_t$
will also form a closed Fermi surface on the top face. So, in this
case, we reach the conclusion that the top face is characterized
also by one Dirac cone.

  The above analysis of the existence of one Dirac cone on the top face has an
essential assumption that in the $k_x=0$ subspace, there is only
  one pair of Kramers's states on the top face at $E$. However,
  according to \texttt{Theorem A}, any odd number
of pairs is possible. Let's consider the case where there are,
say, three pairs $|\psi(k_x=0,\pm k'_{yi})\rangle_t$, $i=1,2,3$
near the $\Gamma$ point of the Brillouin zone for the top face at
the incident energy $E$. Now, there is still perfect transmission
according to Eq.~(3), but the transmitted wave is a linear
combination of three forward-propagating modes on the top face.
For a certain $k_x$ subspace, the boundary conditions for the
scattering spinor wave functions between the top and the front
surfaces are such that the unknown coefficients in the scattering
problem can be solved exactly. This is a general requirement.
Physically, such a boundary condition for continuum wave functions
should be obtained from the underlying lattice system. It is
noteworthy to mention two previous works on such a boundary
condition between regions of qualitatively different
single-particle energy spectrums (the front face and top face are
now qualitatively different in the sense that they have different
number of Dirac cones, see below). First, for graphene/vacuum
boundary, the boundary condition for Dirac particles is nicely
expressed as a constraint of some matrix equation form to the
four-component spinor wave functions\cite{McCann}. Second, for
three types of monolayer/bilayer graphene interfaces, boundary
conditions, i.e., connecting conditions for the continuum wave
functions are obtained from the underlying lattice structure,
based on which the scattering problem of the monolayer/bilayer
graphene interface can be solved\cite{Ando}.

If the top face has multiple pairs of Kramers's states at $k_x=0$
subspace, similar to the single pair case described above, we can
argue that from the continuity principle that the transmitted
waves and their TR partners will form closed Fermi surfaces as the
incident wave $|\psi(k_x,k_{y})\rangle_f$ and the reflected wave
$|\psi(k_x,-k_{y})\rangle_f$ cover the whole Dirac cone of the
front face. Zero transmission probability for some channel will
happen, but only accidentally. The closeness of Fermi surfaces is
a more natural choice(especially for noninteracting system we are
considering).


 \texttt{Theorem C} :\emph{If the low energy spectrum of one surface of a
 TI is
 described by a single Dirac cone, then, the low energy spectrum of any other surfaces should be described
 by an odd number of Dirac cones (though maybe deformed), which are topologically equivalent to the standard Dirac cone.}

In the above, \emph{deformed} means that anisotropy, nonlinearity,
or even particle-hole asymmetry of the dispersion,  are allowed in
general. \texttt{Theorem C} ensures the connectivity property of
low-energy surface states of the 3D TI, which, in combination with
the 2D counterpart given in \texttt{Theorem B}, form the central
connectivity theorems of the edge/surface states of the TIs in
this paper.


 The above discussion is somewhat ideal. We assumed the surface state
can be described by an effective surface Hamiltonian in the bulk
gap region and the boundary between surfaces can be treated as a
geometrical line in the long wave length limit.
Recently, the bulk-surface correspondence in 3D TIs was addressed
by L. Isaev \emph{et al.}\cite{3D bulk-edge correspondence} using
the lattice version of the Dimmock model\cite{franz}. For this
particular model, they found that the number of surface states
intersecting the line connecting two time-reversal-invariant
momenta (i.e., number of deformed Dirac cones for a given sample
surface) can be changed by tuning surface boundary conditions
while the parity of this number remains unchanged. This is in
consistent with our \texttt{Theorem C}. Based on the foregoing
analysis, it is interesting to note that different boundary
conditions used to terminate the 3D lattice at some surface can
change the boundary conditions near the edge between that surface
and other surfaces intersecting with it. When two surfaces are
characterized by different numbers of Dirac cones, the boundary
condition near the edge can be drastically different from the
usual case where there are the same number of Dirac cones for the
intersecting faces. Further model study is needed to explicitly
demonstrate this point.

The connectivity properties embodied in \texttt{Theorem B and C}
are more or less assumed by many researchers. However, as far as
we know, explicit proofs have not been reported so far. For 3D
TIs, such a connectivity property of the surface states for
different surfaces has profound consequences. For example,
transport measurement necessarily involves multiple surfaces
simultaneously. In Ref.~\cite{shunqing shen}, an analysis is given
with respect to the interesting issue of the half conductance
quanta under a magnetic field for Dirac particles living on the 2D
connected surfaces of a 3D TI.

\section{conclusion}
 In conclusion, we have examined the effect of
TRS in the conceptual scattering problem in which helical metals
(whose low-energy excitation is one pair of helical states) are
attached to the central scattering region as electrical leads. The
scattering matrix is found to be antisymmetric so that in any
physically realizable situations, the number of helical metal
leads must be even. Based on this point (\texttt{Theorem A}), we
proved that the quasi-1D helical edge states should always form a
closed loop, thus each edge of a 2D QSH system or any boundary
between 2D topological/nontopological insulators is characterized
by helical states (\texttt{Theorem B}). For the 3D TI, we have
proved that if the low-energy surface states are described by a
single Dirac cone for one surface, then, the low energy excitation
of an arbitrary surface can also be described by an odd number of
Dirac cones (\texttt{Theorem C}) (though they maybe deformed).
These connectivity properties are global properties of TIs. They
result from and are protected by TRS.


\acknowledgments
We acknowledge the financial support from the National
Natural Science Foundation of China (under Grants
No. 11004174 (Y. J. Jiang) and 11174252 (F. Zhai)) and
program for Innovative Research Team in Zhejiang Normal
University. T. Low is partially supported by the NSF Nanoelectronic
Research Initiatives.

\end{document}